\documentclass[aps,prd,preprintnumbers,superscriptaddress,nofootinbib,notitlepage,floatfix]{revtex4-2}
\usepackage[pdftex]{graphicx}
\usepackage{bm,latexsym,amsmath,amssymb,amsfonts,mathrsfs,tikz}
\usetikzlibrary{positioning}
\usepackage{color}
\allowdisplaybreaks[1]
\usepackage[pdftex,colorlinks=true,linkcolor=blue,citecolor=cyan]{hyperref}
\newcommand*{\D}{\mathrm{d}}
\newcommand*{\mpl}{M_{\mathrm{Pl}}}

\begin{document}
\title{Testing gravity with the cosmic microwave background:
constraints on modified gravity with two tensorial degrees of freedom}
%
\author{Takashi~Hiramatsu}
\email[Email: ]{hiramatz@rikkyo.ac.jp}
\affiliation{Department of Physics, Rikkyo University, Toshima, Tokyo 171-8501, Japan
}
\author{Tsutomu~Kobayashi}
\email[Email: ]{tsutomu@rikkyo.ac.jp}
\affiliation{Department of Physics, Rikkyo University, Toshima, Tokyo 171-8501, Japan
}
%
\begin{abstract}
We provide a cosmological test of modified gravity
with two tensorial degrees of freedom and no extra propagating scalar mode.
The theory of gravity we consider admits
a cosmological model that is indistinguishable from the $\Lambda$CDM model
at the level of the background evolution. The model has a single modified-gravity
parameter $\beta$, the effect of which can be seen in linear perturbations,
though no extra scalar mode is propagating. 
Using the Boltzmann code modified to incorporate the present model,
we derive the constraints
$-0.047 < \beta < -0.028$ at 68$\%$ confidence from Planck CMB data.
Since our modified gravity model can hardly be constrained
by the Solar System tests and gravitational-wave propagation,
our result offers the first observational test on the model.
\end{abstract}
\preprint{RUP-22-10}
\maketitle
\section{Introduction}

The cause of the accelerated expansion of the current Universe is
still unknown. Motivated by this mystery, numerous studies have been
made on modified theories of gravity which might be responsible for
the cosmic acceleration.
Modified gravity also plays an essential role in testing general relativity,
as having predictions in modified gravity allows us
to make a comparison among different theories and
confront them with observational data.

To modify general relativity (GR), new dynamical degrees of freedom
are often added on top of two tensorial degrees of freedom corresponding
to two polarization modes of gravitational waves.
In particular, scalar-tensor theories, i.e. modified gravity
with one scalar and two tensorial degrees of freedom, have been studied
extensively, with emphasis on the Horndeski theory~\cite{Horndeski:1974wa, Deffayet:2011gz,Kobayashi:2011nu}
and its extensions~\cite{Langlois:2015cwa,Crisostomi:2016czh,BenAchour:2016fzp} in the past decade (see Ref.~\cite{Kobayashi:2020wqy} for review).
Another way of modifying GR is assuming less symmetries than full diffeomorphism invariance.
Though apparently different, these two ways of modification are basically equivalent.
Consider, for example, a theory of gravity (with no additional fields other than the metric)
having invariance only under
spatial coordinate transformation, $\Vec{x}\to \Vec{x}'=\Vec{x}'(t,\Vec{x})$.
In such a theory, four-dimensional diffeomorphism invariance can be restored by means of
the St\"{u}ckelberg trick, resulting in a fully covariant scalar-tensor theory.
Therefore, a spatially covariant theory of gravity can be regarded as
a gauge-fixed version of a fully covariant scalar-tensor theory.
This idea was used to develop a general framework of scalar-tensor theories~\cite{Gao:2014soa,Gao:2014fra}.
The same idea underlies the construction of the effective field theory
of single-field inflation~\cite{Cheung:2007st}.

It is interesting to note, however, that a spatially covariant theory of gravity
does not always have one scalar and two tensorial degrees of freedom.
Gao and Yao explored spatially covariant theories of gravity
having just two tensorial degrees of freedom and no propagating scalar mode~\cite{Gao:2019twq}.
A well-known example of this kind of theories is the cuscuton theory,
i.e. a k-essence theory with an infinite sound speed~\cite{Afshordi:2006ad}.
The cuscuton theory was extended recently
in Ref.~\cite{Iyonaga:2018vnu} (see
Refs.~\cite{Afshordi:2014qaa,Iyonaga:2020bmm,Panpanich:2021lsd,Maeda:2022ozc,Miranda:2022brj} for
various aspects of extended cuscuton theories).
A family of theories obtained in Ref.~\cite{Gao:2019twq}
may be considered as a further generalization of the cuscuton theory and its extended version.
Phenomenological aspects of modified gravity of~\cite{Gao:2019twq}
were discussed in detail in Ref.~\cite{Iyonaga:2021yfv},
and there it was pointed out that a certain subset of the theories can hardly be
distinguished from GR.
See Ref.~\cite{Bartolo:2021wpt} for inflationary cosmology in
the same modified gravity theory.
Similar modified gravity theories with just two tensorial degrees of
freedom have also been under active discussion~\cite{Lin:2017oow,Carballo-Rubio:2018czn,Aoki:2018zcv,Aoki:2018brq,Mukohyama:2019unx,Feng:2019dwu,DeFelice:2020eju,Aoki:2020oqc,Tasinato:2020fni,DeFelice:2020onz,Yao:2020tur,DeFelice:2020prd,Sangtawee:2021mhz,DeFelice:2021xps,Ganz:2022iiv,DeFelice:2022uxv}.

In this paper, we further explore the cosmology of the cuscuton-like theories of~\cite{Gao:2019twq}.
As elaborated in Ref.~\cite{Iyonaga:2021yfv},
a family of modified gravity theories in~\cite{Gao:2019twq} admits a model
that can mimic the standard cosmic expansion history while evading
basic tests of general relativity such as light bending and
the propagation speed of gravitational waves. 
It is expected, however, that even in such a model
some differences from the standard predictions emerge at the level of linear cosmological perturbations.
The goal of this paper is to clarify how this occurs and
derive constraints on the model parameters from the Planck CMB data.
To do so, we modify the Boltzmann code for scalar-tensor cosmology
developed in~\cite{Hiramatsu:2020fcd}
to be able to handle ``scalarless'' modified gravity models.

This paper is organized as follows.
In the next section, we present the theory of modified gravity
we study in this paper. The action contains three time-dependent
functions which can in principle be chosen freely.
In Sec.~III, we introduce the specific model
that mimics the background evolution of the $\Lambda$CDM model
and is characterized by a single parameter $\beta$.
We then derive the equations governing linear cosmological perturbations
and give some analytic results for the superhorizon and subhorizon limits
in Sec.~IV. Section~V is devoted to the model parameter constraints using our Boltzmann solver and the Markov-Chain Monte-Carlo method with the Planck 2018 likelihood.
We draw our conclusions in Sec.~VI.

\section{A viable theory of spatially covariant gravity with two tensorial degrees of freedom}

To write the action of spatially covariant gravity,
we introduce the ADM variables and use the $(3+1)$ form of the metric,
\begin{align}
    \D s^2=-N^2\D t^2+\gamma_{ij}\left(\D x^i+N^i\D t\right)
    \left(\D x^j+N^j\D t\right),
\end{align}
where $N$ is the lapse function, $N^i$ is the shift vector, and 
$\gamma_{ij}$ is the spatial metric.
The theory of modified gravity we consider in this paper is described by the action
\begin{align}
S=\frac{\mpl^2}{2}\int\D t\D^3x\, \sqrt{\gamma}N\left[
K_{ij}K^{ij}-\frac{1}{3}
\left(\frac{2N}{\beta+N}+1\right)K^2
+R+\alpha_1+\frac{\alpha_3}{N}
\right]+S_{\mathrm{m}},\label{action:Gao2}
\end{align}
where $R$ is the three-dimensional Ricci scalar and 
\begin{align}
    K_{ij}:=\frac{1}{2N}\left(\partial_t\gamma_{ij}-D_iN_j-D_jN_i\right)
\end{align}
is the extrinsic curvature
with $D_i$ being the three-dimensional covariant derivative.
We have the three free functions of $t$ characterizing the theory:
$\beta$, $\alpha_1$, and $\alpha_3$.
We assume that the matter fields have a diffeomorphism invariant action $S_{\textrm{m}}$
and are minimally coupled to the metric.\footnote{Since the gravitational part
of the action is already not diffeomorphism invariant, one might
also assume that diffeomorphism invariance is not maintained
in the action of the matter fields. The cosmological consequences
would then be dependent on the details of the matter action.
To focus on modified gravity effects, we simply assume that
the matter fields have a diffeomorphism invariant action.}

A family of spatially covariant theories of gravity with two tensorial degrees of freedom
and with no extra scalar mode has been developed in Ref.~\cite{Gao:2019twq},
and the gravitational part of the action~\eqref{action:Gao2}
describes a subset of the theories in which no deviation from GR is found
in weak gravitational fields on small scales and gravitational-wave
propagation~\cite{Iyonaga:2021yfv}.

The action~\eqref{action:Gao2} is invariant under a spatial coordinate transformation,
but we no longer have the freedom to perform a temporal coordinate transformation.
Nevertheless, one can always restore the full four-dimensional diffeomorphism invariance
by introducing a St\"{u}ckelberg scalar field. The point is that
the St\"{u}ckelberg field does not propagate if the action takes the above particular form
as shown in Ref.~\cite{Gao:2019twq}.
See Ref.~\cite{Iyonaga:2021yfv} for the fully covariant form of the action
with the St\"{u}ckelberg field.

The $t$-dependent function $\alpha_1(t)$ is essentially the potential for
the St\"{u}ckelberg field (say, $V(\phi)$)
expressed in the unitary gauge ($\phi=\phi(t)$), while
the term $\alpha_3(t)/N$ comes from the square root of its kinetic term,
$\sqrt{-(\partial\phi)^2}= \dot\phi/N$. These two terms constitute
the original version of the cuscuton theory~\cite{Afshordi:2006ad}.
In this paper, we are mostly interested in the novel modification arising
from $\beta$, which was introduced for the first time in Ref.~\cite{Gao:2019twq}
without changing the essential property of the cuscuton theory,
i.e. the number of the dynamical degrees of freedom.
Basically, the impact of $\beta$ shows up only in a cosmological setup~\cite{Iyonaga:2021yfv}, and this term modifies cosmology
both at the levels of the background and perturbation evolution in general.
As described in the next section, however, we can tune the two of the
three functions so that the background evolution is indistinguishable from
that of the standard $\Lambda$CDM model, while one still retains a single
modified gravity parameter to be constrained through linear cosmological perturbations.

Let us present the gravitational field equations derived from the action~\eqref{action:Gao2}.
The Hamiltonian and momentum constraints are given respectively by
\begin{align}
K_{ij}K^{ij}-\frac{1}{3}\left[2\left(\frac{N}{\beta+N}\right)^2+1\right]K^2
-\alpha_1-R
& =-
2\mpl^{-2}\sum_I E^{(I)},\label{Hamilton-C}
\\
D_j\pi_i^j&=-\mpl^{-2}\sum_I J^{(I)}_i,\label{Momentum-C}
\end{align}
while the evolution equations are given by
\begin{align}
&\frac{1}{N\sqrt{\gamma}}\partial_t\left(\sqrt{\gamma}\pi^{kl}\right)
\gamma_{ik}\gamma_{jl}
-\frac{1}{2}\left(\alpha_1+\frac{\alpha_3}{N}\right)\gamma_{ij}
+2\left[
K_{ik}K_j^{\;k}-\frac{1}{3}\left(\frac{2N}{\beta+N}+1\right)KK_{ij}
\right]
\notag \\ &
-\frac{1}{2}\left[
K_{kl}K^{kl}-\frac{1}{3}\left(\frac{2N}{\beta+N}+1\right)K^2
\right]\gamma_{ij}
+\left(R_{ij}-\frac{1}{2}R\gamma_{ij}\right)
+\frac{1}{N}\left(D^2N\gamma_{ij}-D_iD_jN\right)
\notag \\ &
+\frac{1}{N}\left[
D^k(\pi_{ik}N_j)+D^k(\pi_{jk}N_i)-D^k(\pi_{ij}N_k)
\right]
=\mpl^{-2}\sum_IT_{ij}^{(I)},\label{Evolution-eq}
\end{align}
where
\begin{align}
  E^{(I)}&:=(\rho_I+p_I)(Nu^0_I)^2-p_I,
  \\
  J_i^{(I)}&:=(\rho_I+p_I)Nu_I^0 u_{iI},
  \\
  T_{ij}^{(I)}&:=(\rho_I+p_I)u_{iI}u_{jI}+p_I\gamma_{ij}+\Sigma_{ijI},
\end{align}
for each matter component and we defined
\begin{align}
    \pi^{ij}&:=
  K^{ij}-\frac{1}{3}\left(\frac{2N}{\beta+N}+1\right)K\gamma^{ij}.
\end{align}
Here, $\rho_I$, $p_I$, $u^\mu_I$, and $\Sigma_{ijI}$
represent the energy density, the isotropic pressure, the four-velocity, and the
anisotropic stress, respectively, and we consider
baryons ($I=b$), cold dark matter ($I=c$), photons ($I=\gamma$), and neutrinos ($I=\nu$).

\section{Homogeneous and isotropic background}\label{sec:bg}

Now let us consider a homogeneous and isotropic cosmological background.
For the spatially flat model,
the ADM variables are given by
\begin{align}
    N=\bar N(t),\quad N_i=0,\quad \gamma_{ij}=a^2(t)\delta_{ij}.
\end{align}
Substituting these variables to the Hamiltonian constraint and
the evolution equations, we obtain~\cite{Iyonaga:2021yfv}
\begin{align}
    \frac{3H^2}{(\beta/\bar N+1)^2}+\frac{\alpha_1}{2}&=\frac{\rho}{\mpl^2},\label{cosmo:H}
    \\
    -\frac{3H^2}{\beta/\bar N+1}-\frac{2}{\bar N}
    \frac{\D}{\D t}\left(\frac{H}{\beta/\bar N+1}\right)
    -\frac{1}{2}\left(\alpha_1+\frac{\alpha_3}{\bar N}\right)&=\frac{p}{\mpl^2},\label{cosmo:E}
\end{align}
where
$H:=\bar N^{-1}\D \ln a/\D t$
is the Hubble parameter and we defined
$\rho=\sum_I\rho_I$ and $p=\sum_I p_I$.
Since the total energy-momentum tensor is assumed to be covariantly conserved,
we have
\begin{align}
    \bar N^{-1}\dot \rho+3H(\rho+p)=0,\label{cosmo:cons}
\end{align}
where a dot stands for differentiation with respect to $t$.

In general relativity, the conservation equation~\eqref{cosmo:cons}
can be derived from the Hamiltonian constraint and the evolution equations,
and hence it does not yield an independent equation.
In the present case, however, the conservation equation~\eqref{cosmo:cons} is
independent of Eqs.~\eqref{cosmo:H} and~\eqref{cosmo:E}.
Combining Eqs.~\eqref{cosmo:H}--\eqref{cosmo:cons}, we obtain
\begin{align}
    \dot\alpha_1-\sqrt{3}\left(\frac{\rho}{\mpl^2}-\frac{\alpha_1}{2}\right)^{1/2}
    \left[\alpha_3-\left(\alpha_1+\frac{2p}{\mpl^2}\right)\beta\right]=0,
    \label{cosmo:cusuc}
\end{align}
which we will use below instead of Eq.~\eqref{cosmo:E}.
In the covariant formulation, this equation corresponds to
the equation of motion for the St\"{u}ckelberg field.


In this paper, we are interested in the background solution
which is identical to that in GR:
$\bar N=1$, $H=H_{\mathrm{GR}}(t)$.\footnote{Note that
we do not have the freedom to choose the temporal coordinate so that $\bar N=1$.
For given $\alpha_1(t)$, $\alpha_3(t)$, and $\beta(t)$,
the lapse function is determined as a solution to
Eqs.~\eqref{cosmo:H},~\eqref{cosmo:cons}, and~\eqref{cosmo:cusuc}.
Having said that, for simplicity, we consider a theory that admits $\bar N=1$ as a solution.}
We therefore design the time dependence
of $\alpha_1$, $\alpha_3$, and $\beta$ so that the theory admits
such a solution. This can be done as follows.
Since $H_{\mathrm{GR}}(t)$ (with $\bar N=1$) is a solution to the Einstein equations with a
cosmological constant $\Lambda$,
it satisfies
\begin{align}
    3H_{\mathrm{GR}}^2&=\frac{\rho}{\mpl^2}+\Lambda,\label{GRHam}
    \\
    -2\dot H_{\mathrm{GR}}&=\frac{\rho+p}{\mpl^2}.\label{GREv}
\end{align}
If $H=H_{\mathrm{GR}}(t)$ is a solution to Eqs.~\eqref{cosmo:H} and~\eqref{cosmo:E},
then $\alpha_1$ and $\alpha_3$ must satisfy, for any $\beta=\beta(t)$,
\begin{align}
    \alpha_1(t)&=\frac{6\beta(2+\beta)}{(1+\beta)^2}H_{\mathrm{GR}}^2-2\Lambda,
    \\
    \alpha_3(t)&=4\frac{\D}{\D t}\left(\frac{\beta H_{\mathrm{GR}}}{1+\beta}\right)
    -\frac{6\beta H_{\mathrm{GR}}^2}{(1+\beta)^2}.
\end{align}
Thus, in the theory with $\alpha_1$, $\alpha_3$, and $\beta$ satisfying these two equations,
the solution in GR, $H=H_{\mathrm{GR}}$, can be reproduced.
In this paper, we consider the simplest case with
\begin{align}
    \beta=\mathrm{const},
\end{align}
and study CMB constraints on the parameter $\beta$
that controls the deviation from GR at linear perturbation order.

\section{Cosmological perturbations}

\subsection{Basic equations}

In what follows we consider the background with
$\bar N=1$ as in the background model introduced in the previous section.
The perturbed ADM variables are given by
\begin{align}
    N=1+\delta n,\quad N_i=a \partial_i\chi,
    \quad \gamma_{ij}=a^2(1-2\psi)\delta_{ij},
\end{align}
where we used the spatial gauge degrees of freedom
to write $\gamma_{ij}$ in the above form.
The perturbed four-velocity of each matter component is written as
\begin{align}
    u^0_I=1-\delta n,\quad u_{I}^i=a^{-1}\partial^i\widetilde v_I,
\end{align}
while the energy density, the isotropic pressure, and the anisotropic stress are expressed as
\begin{align}
    \rho_I=\bar\rho_I(1+\widetilde \delta_I), \quad p_I=0, \quad \Sigma_{ijI}=0,
\end{align}
for $I=b,c$ and 
\begin{align}
    \rho_I=\bar\rho_I(1+\widetilde \delta_I), \quad p_I=\frac{1}{3}\rho_I,
   \quad
    \Sigma_{ijI}=p_I\left(\partial_i\partial_j-\frac{1}{3}\delta_{ij}\partial^2\right)
    \widetilde{\delta\Pi}_I,
\end{align}
for $I=\gamma,\nu$. Here, we put the tilde to emphasize that
$\widetilde \delta_I$, $\widetilde v_I$, and $\widetilde{\delta\Pi}_I$ are
defined in the unitary gauge in which the action is written as Eq.~\eqref{action:Gao2}.

The gravitational field equations~\eqref{Hamilton-C}--\eqref{Evolution-eq}
in the Fourier space yield
\begin{align}
    &\frac{3H}{(1+\beta)^2}\dot\psi+\frac{3H^2}{(1+\beta)^3}\delta n 
    -\frac{H}{(1+\beta)^2}\frac{k^2}{a}\chi+\frac{k^2}{a^2}\psi 
    =-\frac{\widetilde{\delta\rho}}{2\mpl^2},\label{pert:Ham}
    \\
    &\frac{\dot\psi}{1+\beta}+\frac{H}{(1+\beta)^2}\delta n +\frac{\beta}{3(1+\beta)} 
    \frac{k^2}{a}\chi=-\frac{\widetilde{\delta q}}{2\mpl^2},\label{pert:Mom}
    \\
    &\partial_t\left(\frac{\dot\psi}{1+\beta}\right)+\frac{3H}{1+\beta} \dot\psi
    +\frac{H}{(1+\beta)^2}\dot{\delta n}
    +
    \left\{
    \partial_t\left[\frac{(2+\beta)H}{(1+\beta)^2}\right]
    +\frac{3(2+\beta)H^2}{2(1+\beta)^2}
    \right\}\delta n
    \notag \\ & +\frac{k^2}{3a^2}
    \left[
    \psi-\delta n-\partial_t\left(\frac{a\chi}{1+\beta}\right)-\frac{aH\chi}{1+\beta} 
    \right]=\frac{\widetilde{\delta p}}{2\mpl^2},\label{pert:Trace}
    \\
    &\psi-\delta n-\partial_t(a\chi)-aH\chi=\frac{a^2\widetilde{\delta\Sigma}}{\mpl^2},
    \label{pert:Traceless}
\end{align}
where $k$ is the wave number and
\begin{align}
    \widetilde{\delta\rho}&:=\sum_I\rho_I\widetilde\delta_I,
    \\
    \widetilde{\delta q}&:=a \sum_I(\rho_I+p_I)(\widetilde v_I+\chi),
    \\
    \widetilde{\delta p}&:=\frac{1}{3}\left(\rho_\gamma\widetilde
    \delta_\gamma+\rho_\nu\widetilde\delta_\nu\right),
    \\
    \widetilde{\delta \Sigma} &:=p_\gamma\widetilde{\delta\Pi}_\gamma
    +p_\nu\widetilde{\delta\Pi}_\nu. 
\end{align}
Note that the evolution equations can be split into
the trace and traceless parts, and the former gives Eq.~\eqref{pert:Trace},
while the latter reduces to Eq.~\eqref{pert:Traceless}.
Since the total energy-momentum tensor is assumed to be covariantly conserved, we have 
\begin{align}
    \dot{\widetilde{\delta\rho}}+3H\widetilde{\delta\rho}-3\sum_I(\rho_I+p_I)\dot\psi
    &=
    \frac{k^2}{a^2}\biggl[\widetilde{\delta q}-a\sum_I(\rho_I+p_I)\chi\biggr],
    \label{conseq:0}
    \\
    \dot{\widetilde{\delta q}}+3H\widetilde{\delta q} +\sum_I(\rho_I+p_I)\delta n
    +\widetilde{\delta p }
    &=\frac{2k^2}{3}\widetilde{\delta\Sigma}.\label{conseq:i}
\end{align}

In GR, Eq.~\eqref{conseq:0} is an automatic consequence
of Eqs.~\eqref{pert:Ham},~\eqref{pert:Mom}, and \eqref{pert:Trace}.
In the present case, however, Eq.~\eqref{conseq:0} is an independent equation.
Combining
Eqs.~\eqref{pert:Ham},~\eqref{pert:Mom},~\eqref{pert:Trace} and~\eqref{conseq:0}
and using the background equations,
we obtain
\begin{align}
    \mpl^2\frac{k^2}{a}\chi&=
    \frac{2\beta(1+\beta)H}{B_0}\left(\widetilde{\delta\rho}+3\widetilde{\delta p}\right)
    -\frac{3}{2}\widetilde{\delta q},\label{constchi}
\end{align}
where
\begin{align}
    B_0&:=2\beta\left[(5+3\beta)H^2+2(1+\beta)\dot H-2\dot \beta H\right]
    -\frac{4}{3}\beta(1+\beta)^2\frac{k^2}{a^2}+(1+\beta)^3\alpha_3.
\end{align}
This equation algebraically determines $\chi$ in terms of
$\widetilde\delta_I$ and $\widetilde v_I$.
Note that Eq.~\eqref{conseq:i} is not an independent equation.

It is more convenient to express the equations in terms of
the variables in the Newtonian gauge, as they are
used in the numerical code.
Under a first-order change from the unitary gauge,
\begin{align}
    t'=t+\xi(t,\Vec{x}),
\end{align}
the first-order perturbations transform as
\begin{align}
    &\delta n'=\delta n-\dot\xi,\quad \chi'=\chi+a^{-1}\xi,
    \quad \psi'=\psi+H\xi,
    \notag \\
    &\delta_I'=\widetilde\delta_I-\frac{\dot\rho_I}{\rho_I}\xi,
    \quad 
    v_I'=\widetilde v_I,\quad \delta \Pi_I'=\widetilde{\delta \Pi}_I.
\end{align}
Here $\xi$ is the fluctuation of the St\"{u}ckelberg field in the new coordinate system.
These transformation rules lead us to introduce the
gravitational potentials in the Newtonian gauge defined as
\begin{align}
    \Phi:=\delta n+\partial_t\left(a\chi\right),
    \quad 
    \Psi:=\psi-aH\chi ,
\end{align}
and the matter quantities in the Newtonian gauge defined as
\begin{align}
    \delta_I:=\widetilde \delta_I+\frac{\dot\rho_I}{\rho_I}a\chi,
    \quad 
    v_I:=\widetilde v_I,\quad \delta\Pi_I:=\widetilde{\delta\Pi}_I.
\end{align}
Using these variables,
the Hamiltonian and momentum constraints can be written, respectively, as
\begin{align}
    &\frac{3{\cal H}}{(1+\beta)^2}\Psi'+\frac{3{\cal H}^2}{(1+\beta)^3}\Phi+k^2\Psi
    +\frac{3\beta{\cal H}^2}{(1+\beta)^3}\chi'
    +\frac{\beta(2+\beta)}{(1+\beta)^2}k^2{\cal H}\chi
    \notag \\ & 
    +\frac{3{\cal H}}{4(1+\beta)^2}\left[
    \frac{2\beta(1-\beta)}{1+\beta}{\cal H}^2
    -4\beta{\cal H}'+4\beta'{\cal H}
    -a^2(1+\beta)^2\alpha_3
    \right]\chi
    \notag \\ & 
    =-\frac{a^2}{2\mpl^2}\sum_I\rho_I\delta_I,\label{N:conf:Ham}
    \\
    &\frac{\Psi'}{1+\beta}+\frac{{\cal H}\Phi}{(1+\beta)^2}+
    \frac{\beta{\cal H}}{(1+\beta)^2}\chi'+\frac{\beta}{3(1+\beta)}k^2\chi 
    -\left[\frac{2\beta{\cal H}^2-4\beta'{\cal H}+a^2(1+\beta)^2\alpha_3}{4(1+\beta)^2}\right]\chi 
    \notag \\ &
    =-\frac{a^2}{2\mpl^2}\sum_I(\rho_I+p_I) v_I,\label{N:conf:Mom}
\end{align}
while the traceless part of the evolution equations is expressed as
\begin{align}
    \Psi-\Phi=\frac{a^2}{\mpl^2}\left(p_\gamma\delta\Pi_\gamma+p_\nu\delta\Pi_\nu\right),
    \label{N:conf:traceless}
\end{align}
where a dash stands for differentiation with respect to
the conformal time defined by $\D\eta=\D t/a$ and ${\cal H}:=a'/a$.
It should be emphasized that the traceless part equation~\eqref{N:conf:traceless}
is exactly the same as the corresponding equation in GR.
These equations must be supplemented with the
constraint equation~\eqref{constchi} for $\chi$, which can now be written as
\begin{align}
    \mpl^2\chi=\frac{2\beta(1+\beta)^3{\cal H}}{B_1}\cdot a^2\sum_I(\rho_I+3p_I)\delta_I+
    \frac{B_2}{B_1}\cdot a^2\sum_I(\rho_I+p_I)v_I,
    \label{N:conf:chi}
\end{align}
where 
\begin{align}
B_1&=\frac{3}{4}\left\{
2(2-\beta){\cal H}^2+4\beta'{\cal H}-(1+\beta)\left[
4{\cal H}'+ a^2(1+\beta)\alpha_3
\right]
\right\}
\notag \\
&\quad \times
\left\{
2\beta\left[
(1-\beta-6c_s^2(1+\beta)){\cal H}^2+2(1+\beta){\cal H}'-2\beta'{\cal H}
\right]+a^2(1+\beta)^3\alpha_3
\right\}
\notag \\ &\quad
+ \left\{
2\beta(1+\beta)^2
\left[(1+2\beta){\cal H}^2+4(1+\beta){\cal H}'-4\beta'{\cal H}\right]
+a^2(1+\beta)^4(1+2\beta)\alpha_3
\right\}k^2
-\frac{4}{3}\beta(1+\beta)^4k^4,
\\
B_2&=
-\frac{3}{2}a^2\alpha_3(1+\beta)^5+\beta(1+\beta)^2
\left[
2k^2(1+\beta)^2-3(3+\beta){\cal H}^2-6(1+\beta){\cal H}'+6\beta'{\cal H}
\right],
\end{align}
with $c_s^2:=\sum_Ip_I'/\sum_I\rho_I'$.
Now $\chi$ may be regarded as a fluctuation of the St\"{u}ckelberg field
in the Newtonian gauge. However, $\chi$ is not dynamical
because it is determined from Eq.~\eqref{N:conf:chi} which is not of the form of
a hyperbolic evolution equation.

\subsection{Initial conditions}

Having obtained the perturbed gravitational field equations,
let us discuss the initial conditions for perturbations set deep in the radiation era,
where the scale factor is given by $a\propto \eta$ and hence ${\cal H}=\eta^{-1}$.
Since we are mainly interested in the background model introduced in the previous section,
we assume that $a^2\alpha_1\propto {\cal H}^2$, $a^2\alpha_3\propto {\cal H}^2$,
and $\beta=\,$constant deep in the radiation era.

By a direct manipulation one can see that our equations admit
the following solution in the early time during radiation domination:
\begin{align}
    &\Phi=2C_1,\quad \Psi=2C_1\left(1+\frac{2f_\nu}{5}\right),\label{init01}
    \\
    &\delta_\gamma=\delta_\nu=\frac{4}{3}\delta_b=\frac{4}{3}\delta_c=-4C_1\label{init02},
    \\
    &v_\gamma=v_\nu=v_b=v_c=-\chi=-C_1\eta\approx 0,\label{init03}
\end{align}
where $C_1$ is a constant and $f_\nu:=\rho_\nu/(\rho_\gamma+\rho_\nu)$.
We have taken into account the neutrino quadrupole moment,
which is reflected in the relation between $\Phi$ and $\Psi$.
The relation can be derived by using the Boltzmann equation for neutrinos and
the traceless part of the evolution equations, which remain the same as in GR.
The constant $C_1$ is related to the comoving curvature perturbation
(which turns out to be constant and coincide with
the curvature perturbation on uniform-density slices for $k\eta\ll 1$),
\begin{align}
    \zeta&=-\psi+\frac{H\widetilde{\delta q}}{\sum_I(\rho_I+p_I)}
    \notag \\ &
    =-\Psi+\frac{{\cal H}\sum_I(\rho_I+p_I)v_I}{\rho+p},
\end{align}
as 
\begin{align}
    C_1=-\frac{5}{4f_\nu+15}\zeta.
\end{align}
We have thus set up the initial conditions in the early time during radiation domination.
Note that the initial conditions~\eqref{init01} and~\eqref{init02}
are the same as those used in the standard cosmological setup in GR.
This comes with surprise because the perturbed
gravitational field equations except for the
traceless part are apparently different from the conventional
Einstein equations in the radiation era.
It turns out, however, that various modifications cancel out
in the early time during radiation domination.

\subsection{Late-time behavior}

We then move to study the late-time evolution of perturbation modes
for which $k/{\cal H}\gg 1$ is satisfied.
From Eq.~\eqref{N:conf:chi}, we have, for $k\gg {\cal H}$,
\begin{align}
    k^2{\cal H}\chi \approx
    -\frac{3}{2(1+\beta)}\cdot \frac{{\cal H}^2}{k^2}\cdot 
    \frac{a^2}{\mpl^2}\sum_I(\rho_I+3p_I)\delta_I
    -\frac{3}{2\beta(1+\beta)^2}\cdot \frac{a^2{\cal H}}{\mpl^2}\sum_I 
    (\rho_I+p_I)v_I.
\end{align}
This equation, together with Eqs.~\eqref{N:conf:Ham} and~\eqref{N:conf:Mom},
yields
\begin{align}
    k^2\Psi\approx -\frac{a^2}{\mpl^2}\sum_I\rho_I\delta_I 
    \gg {\cal H}^2\Phi\sim \beta k^2{\cal H}\chi\sim 
     \frac{a^2{\cal H}}{\mpl^2}\sum_I (\rho_I+p_I)v_I,
\end{align}
and $v_I\gg\chi$ for $k\gg{\cal H}$. Therefore,
in the subhorizon regime at late times,
the leading part of the Hamiltonian constraint
reduces to the same Poisson equation as in GR.
The traceless part of the gravitational field equations and
the fluid equations are also the same as in GR.
(Since $v_I\gg\chi$ for the subhorizon modes,
the $\chi$'s contribution to the fluid equations is negligible.)
Thus, aside from the overall amplitude, we have the same subhorizon evolution of
the perturbations as in GR.
This fact was shown in the case of a pressureless fluid in Ref.~\cite{Iyonaga:2021yfv},
and here we have generalized their result to the case of a mixture of fluids.

Having the above arguments in mind, let us consider, for example,
the evolution of the gravitational potential $\Phi$ normalized by the
solution in GR: $\Phi/\Phi_{\textrm{GR}}$.
Suppose that we impose the initial condition so that $\Phi/\Phi_{\textrm{GR}}=1$ at $\eta=0$.
Modified gravity effects come into play on intermediate scales, $k\sim {\cal H}$,
and $\Phi/\Phi_{\textrm{GR}}$ then evolves with time, deviating from 1.
At late times, $k\gg{\cal H}$, $\Phi/\Phi_{\textrm{GR}}$ settles down to a constant value
which in general differs from 1.
This behavior will modify the predictions of CMB anisotropies,
leading to constraints on the model parameter $\beta$,
as elaborated in the next section.

\section{CMB constraints on model parameters}

\subsection{Best-fit parameters}

\begin{table}[t]
\begin{tabular}{c|cc}
\hline\hline
    & 2DoF & $\Lambda$CDM \\ \hline
$\beta$               & $-0.0388_{-0.0083}^{+0.011}$ &                                  \\[0.4em]
$10^{9}A_{{\rm s}}e^{-2\tau}$ & $1.8902_{-0.0062}^{+0.0090}$ & $1.8868_{-0.0056}^{+0.0066}$     \\[0.4em]
$n_{{\rm s}}$                 & $0.9748_{-0.0025}^{+0.0035}$ & $0.9697_{-0.0027}^{+0.0042}$     \\[0.4em]
$h$                   & $0.6853_{-0.0051}^{+0.0027}$ & $0.6782_{-0.0056}^{+0.0038}$     \\[0.4em]
$h^2\Omega_{{\rm c}}$       & $0.11843_{-0.00069}^{+0.0010}$ & $0.11878_{-0.00064}^{+0.0015}$   \\[0.4em]
$h^2\Omega_{{\rm b}}$       & $0.02172_{-0.00012}^{+0.00011}$ & $0.02177_{-0.00013}^{+0.00014}$  \\[0.4em]
$\tau$                & $0.0497_{-0.0052}^{+0.0052}$ & $0.0489_{-0.0045}^{+0.0058}$    \\[0.4em]
\hline
$\ln \mathcal{L}$    & $-1426$ & $-1432$ \\[0.4em]
\hline\hline
\end{tabular}
\caption{1$\sigma$ confidence ranges of the model parameters in the modified gravity model ({\it left}) and the $\Lambda$CDM model ({\it right}). In the bottom line, we show the values of the likelihood for the best-fit parameter set.}
\label{tab:conf_range}
\end{table}

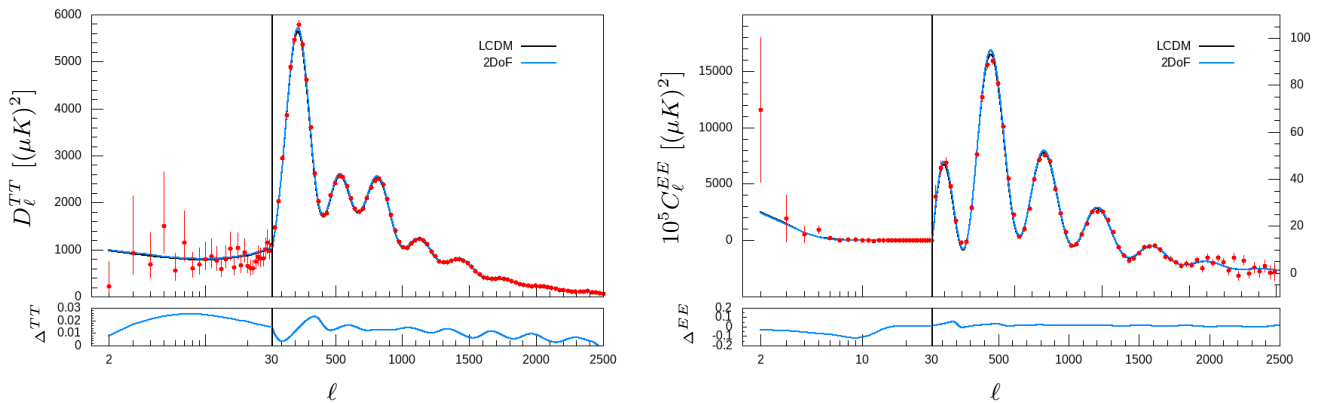
\begin{figure}[t]
\begin{tikzpicture}
\node (img) {\includegraphics[width=8cm]{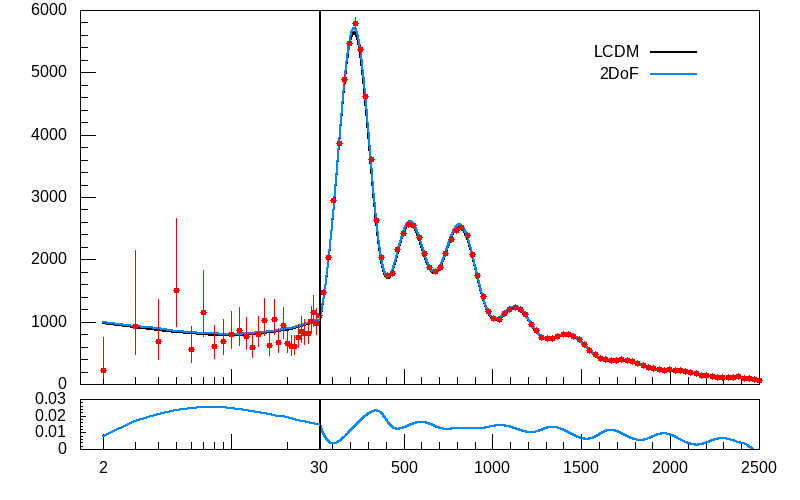}};
\node[below=of img, node distance=0cm, yshift=1.0cm, xshift=0.0cm, anchor=center, font=\small] {$\ell$};
\node[left=of img, node distance=0cm, rotate=90, anchor=center, yshift=-1.0cm, xshift=0.5cm, font=\small] {$D^{TT}_{\ell}~[(\mu K)^2]$};
\node[left=of img, node distance=0cm, rotate=90, anchor=center, yshift=-1.2cm, xshift=-1.7cm, font=\tiny] {$\Delta^{TT}$};
\end{tikzpicture}
\begin{tikzpicture}
\node (img) {\includegraphics[width=8.4cm]{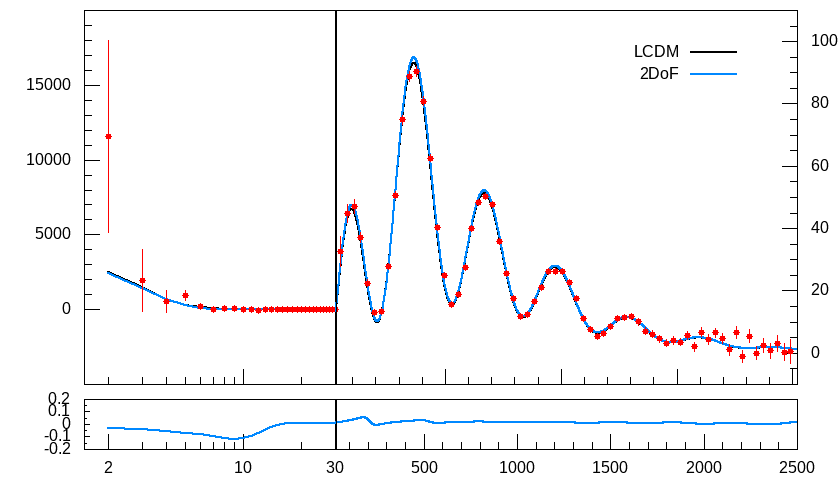}};
\node[below=of img, node distance=0cm, yshift=1.0cm, xshift=0.0cm, anchor=center, font=\small] {$\ell$};
\node[left=of img, node distance=0cm, rotate=90, anchor=center, yshift=-1.0cm, xshift=0.5cm, font=\small] {$10^{5}C^{EE}_{\ell}~[(\mu K)^2]$};
\node[left=of img, node distance=0cm, rotate=90, anchor=center, yshift=-1.2cm, xshift=-1.7cm, font=\tiny] {$\Delta^{EE}$};
\end{tikzpicture}
\caption{Angular power spectra of the temperature fluctuations ({\it left}) and the E-mode polarization ({\it right}) with the best-fit parameters. The elongated panels below the spectra show the fractional deviation from the $\Lambda$CDM case, $\Delta^{X} := (C_\ell^{X({\rm 2DoF})}-C_\ell^{X(\Lambda{\rm CDM})})/C_\ell^{X(\Lambda{\rm CDM})}$ for $X=TT, EE$. Notice that, in the right panel, the vertical scale for $\ell\geq 30$ shown in the right vertical axis is different from that for $\ell < 30$ shown in the left vertical axis. }
\label{fig:bestfit}
\end{figure}

We determine the best-fit parameters and the confidence ranges allowed by the Planck 2018 observations using the 
Markov-Chain Monte-Carlo (MCMC) method with Planck 2018 TTTEEE+lowE likelihood.\footnote{The Planck likelihood files we use here are {\tt commander\_dx12\_v3\_2\_29.clik} for $TT$ in $2\leq \ell\leq 29$, {\tt plik\_rd12\_HM\_v22b\_TTTEEE.clik} for $TT+TE+EE$ in $30\leq \ell\leq 2508$, and {\tt simall\_100x143\_offlike5\_EE\_Aplanck\_B.clik} for $EE$ in $2\leq \ell \leq 29$.}
We use the Boltzmann solver developed in Ref.~\cite{Hiramatsu:2020fcd}, and modify it to implement Eq.~\eqref{N:conf:Ham} in the gravity sector. 
We vary the parameter $\beta$ characterizing the present model in addition to the standard parameters in the $\Lambda$CDM model; the amplitude of the curvature perturbation, $B_{{\rm s}} := A_{{\rm s}}e^{-2\tau}$, the spectral index, $n_{{\rm s}}$, the reduced Hubble parameter, $h$, the fractional amount of the CDM, $h^2\Omega_{\rm c}$, that of the baryons, $h^2\Omega_{\rm b}$ and the optical depth, $\tau$.

In Table \ref{tab:conf_range}, we show the resultant confidence ranges of the seven model parameters obtained by the MCMC simulation. The superscript and subscript indicate the 68\% confidence range, and we also show the maximum value of the logarithmic likelihood, $\ln\mathcal{L}$, in each model. 
It is thus found
that the best-ﬁt range for $\beta$ is
\begin{align}
    -0.047 < \beta < -0.028
\end{align}
at 68$\%$ confidence level, which indicates a $\sim 4\sigma$ deviation from GR.

In Fig.~\ref{fig:bestfit}, we show the rescaled angular power spectra, $D_\ell := \ell(\ell+1)C_\ell/(2\pi)$, for the temperature fluctuations and
$C_\ell$ for the E-mode polarization 
with the best-fit parameters in the present modified gravity
model and the $\Lambda$CDM model. As the differences between them are too small to be seen from the power spectra, we also show the fractional change of $D^{X}_\ell$ or $C^{X}_\ell$ defined as $\Delta^{X} := (C_\ell^{X({\rm 2DoF})}-C_\ell^{X(\Lambda{\rm CDM})})/C_\ell^{X(\Lambda{\rm CDM})}$ for $X=TT, EE$
in the elongated panel below the angular power spectrum.
(The superscript ``2DoF'' is used to denote the angular power
spectrum in the modified gravity model.)
We thus find that the differences are suppressed within a few percent for $D^{TT}_\ell$ and $\sim 10\%$ for $C^{EE}_\ell$. 

\begin{figure}[t]
\begin{tikzpicture}
\node (img) {\includegraphics[width=8cm]{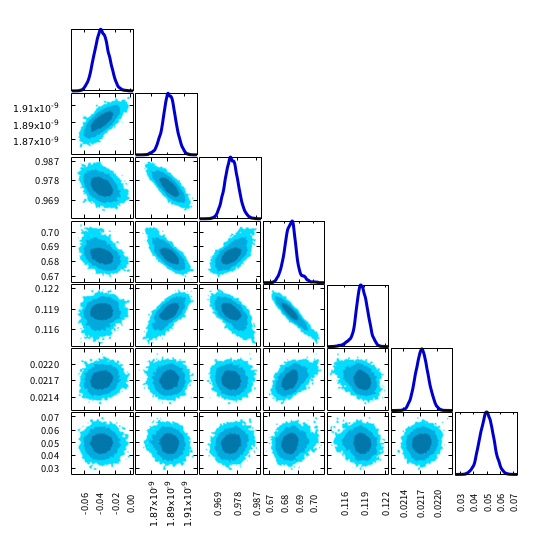}};
\node[below=of img, node distance=0cm, yshift=1.2cm, xshift=-2.7cm, anchor=west, font=\tiny] {$\beta\hspace{6.5mm}B_{{\rm s}}\hspace{5.8mm}n_{{\rm s}}\hspace{6.4mm}h\hspace{5.7mm}h^2\Omega_{\rm c}\hspace{4.0mm}h^2\Omega_{\rm b}\hspace{6.0mm}\tau$};
\node[left=of img, node distance=0cm, rotate=90, anchor=west, yshift=-1.0cm, xshift=-2.6cm, font=\tiny] {$\tau\hspace{5.2mm}h^2\Omega_{\rm b}\hspace{3.5mm}h^2\Omega_{\rm c}\hspace{6.0mm}h\hspace{7.5mm}n_{{\rm s}}\hspace{6.0mm}B_{{\rm s}}\hspace{6.5mm}\beta$};
\end{tikzpicture}
\begin{tikzpicture}
\node (img) {\includegraphics[width=7.2cm]{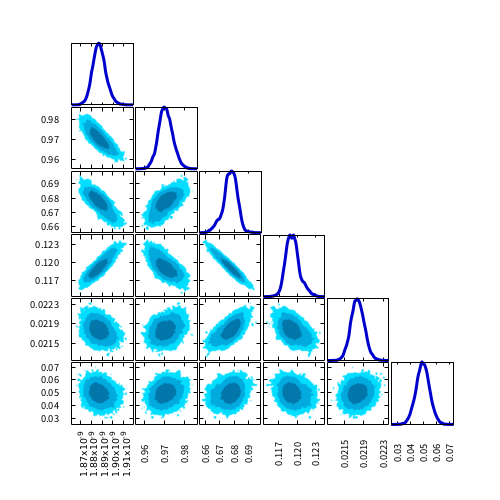}};
\node[below=of img, node distance=0cm, yshift=1.2cm, xshift=-2.3cm, anchor=west, font=\tiny] {$B_{{\rm s}}\hspace{5.8mm}n_{{\rm s}}\hspace{7.0mm}h\hspace{6.0mm}h^2\Omega_{\rm c}\hspace{3.0mm}h^2\Omega_{\rm b}\hspace{5.0mm}\tau$};
\node[left=of img, node distance=0cm, rotate=90, anchor=west, yshift=-1.0cm, xshift=-2.2cm, font=\tiny] {$\tau\hspace{5.2mm}h^2\Omega_{\rm b}\hspace{3.5mm}h^2\Omega_{\rm c}\hspace{5.0mm}h\hspace{7.5mm}n_{{\rm s}}\hspace{6.5mm}B_{{\rm s}}$};
\end{tikzpicture}
\caption{68$\%$, 95$\%$ and 99$\%$ confidence contours in the modified gravity model ({\it left}) and the $\Lambda$CDM model ({\it right}).}
\label{fig:contour}
\end{figure}
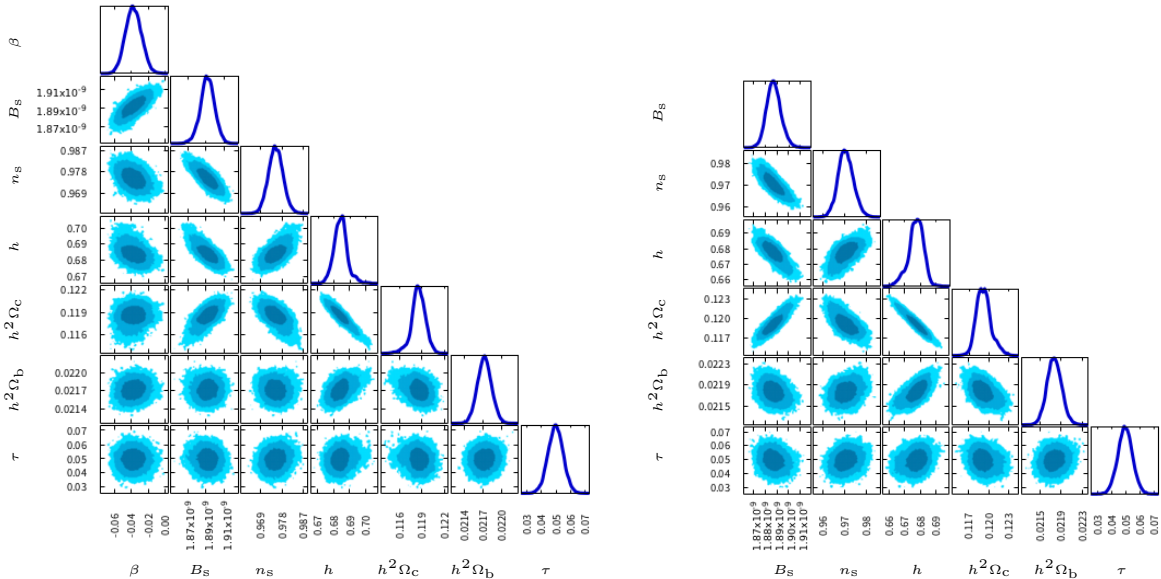

The multivariate distribution function for the model parameters are presented in Fig.~\ref{fig:contour}. 
There is a weak positive correlation between $\beta$ and $B_{{\rm s}}$
and no significant correlations are found with other parameters.
This indicates that the parameter $\beta$ 
can be treated as a parameter controlling the overall amplitude of the anisotropies in a similar manner to the primordial amplitude around the best-fit parameter set. 
As a result, the goodness of the fitting is not significantly improved by introducing the new parameter $\beta$ as shown in Table \ref{tab:conf_range}. The impact of the parameter $\beta$ on the angular power spectrum will be discussed in more detail
in the next subsection.


\subsection{Impact of the $\beta$ parameter}

\begin{figure}[h]
\begin{tikzpicture}
\node (img) {\includegraphics[width=8cm]{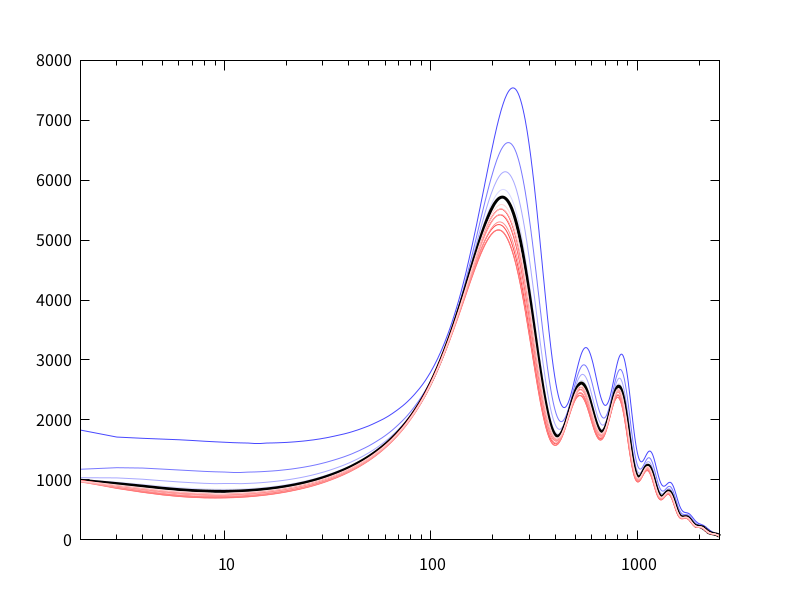}};
\node[below=of img, node distance=0cm, yshift=1.0cm, xshift=0.0cm, anchor=center, font=\small] {$\ell$};
\node[left=of img, node distance=0cm, rotate=90, anchor=center, yshift=-1.0cm, xshift=0.0cm, font=\small] {$D^{TT}_{\ell}~[(\mu K)^2]$};
\end{tikzpicture}
\begin{tikzpicture}
\node (img) {\includegraphics[width=8cm]{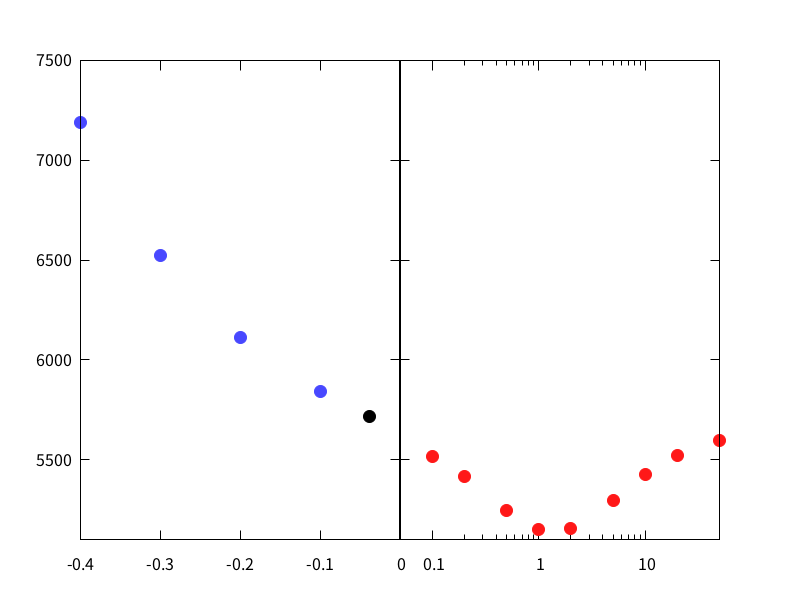}};
\node[below=of img, node distance=0cm, yshift=1.0cm, xshift=0.0cm, anchor=center, font=\small] {$\beta$};
\node[left=of img, node distance=0cm, rotate=90, anchor=center, yshift=-1.0cm, xshift=0.0cm, font=\small] {$D^{TT}_{220}~[(\mu K)^2]$};
\end{tikzpicture}
\caption{({\it Left}) Angular power spectrum of temperature anisotropies, $D^{TT}_\ell := \ell(\ell+1)C^{TT}_\ell/2\pi$, for various $\beta$ in the unit of $(\mu{\rm K})^2$. The black solid line is $D^{TT}_\ell$ for $\beta=-0.0388$, the red lines are for $0.1 \leq \beta\leq 50$ and the blue lines are for $-0.4 \leq \beta \leq -0.1$. ({\it Right}) We also show the value of $D^{TT}_\ell$ at $\ell=220$ for various $\beta$, in which the color pattern is the same used in the left panel. Note that we put together two plots for $\beta\leq 0$ (linear horizontal axis) and $\beta>0$ (logarithmic).}
\label{fig:cl}
\end{figure}
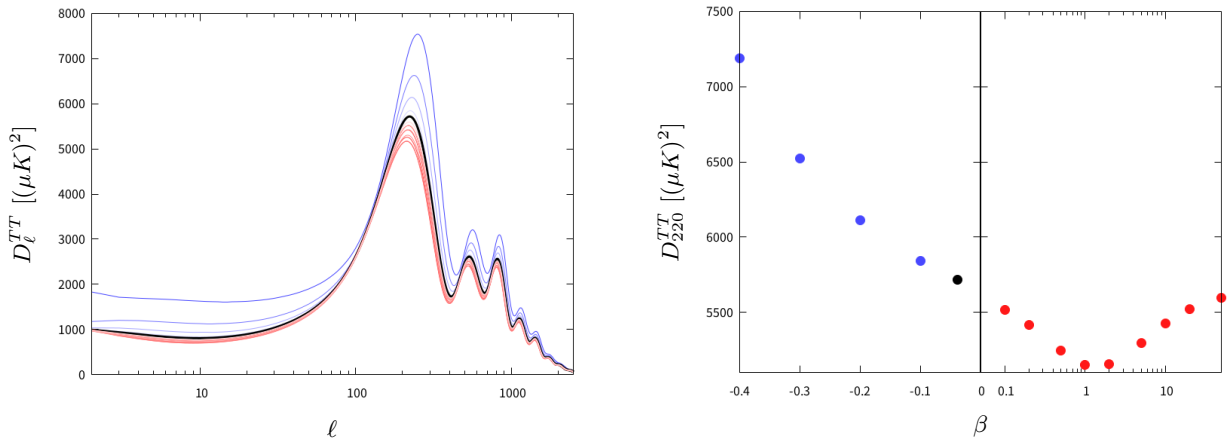

In the left panel of Fig.~\ref{fig:cl}, we present the angular power spectra of temperature anisotropies for various $\beta$. The black line represents the one with the best-fit parameters given in ``2DoF'' in Table \ref{tab:conf_range}. Varying $\beta$ from $-0.1$ to $-0.4$, we obtain the power spectra represented by blue curves and find the monotonic enhancement of the amplitude.
In contrast, varying $\beta$ from 0.1 to 50, we obtain the power spectra represented by red curves whose dependence on $\beta$ is not monotonic. To clarify this, we also show the value of $D_\ell$ at $\ell=220$ for various $\beta$ in the right panel of Fig.~\ref{fig:cl}. As is seen from the figure, the amplitude of the first peak of the angular power spectrum is minimum for $\beta \approx 1.5$, and the amplitude turns to be larger for $\beta > 1.5$ as $\beta$ is increased. 
Therefore, in the vicinity of the best-fit value ($\beta=-0.0388$), 
a large negative value of $\beta$ provides the same effect as
a large value of $B_{\rm s}$. In other words, smaller $|\beta|$ is
preferred for larger $B_{\rm s}$ to compensate each other.
This positive correlation can be found in the small panel for $\beta$ vs $B_{{\rm s}}$ in the left panel of Fig.~\ref{fig:contour}.

\begin{figure}[h]
\begin{tikzpicture}
\node (img) {\includegraphics[width=8cm]{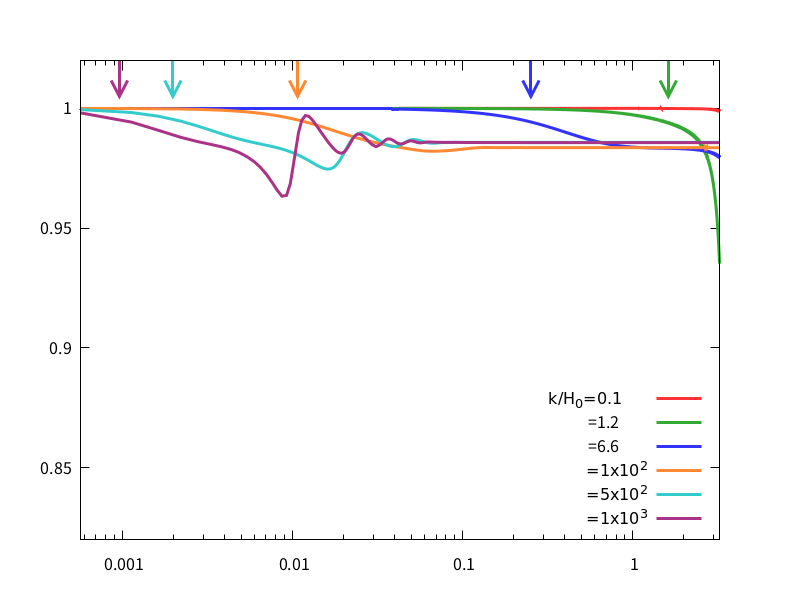}};
\node[below=of img, node distance=0cm, yshift=1.0cm, xshift=0.0cm, anchor=center, font=\small] {$H_0\eta$};
\node[left=of img, node distance=0cm, rotate=90, anchor=center, yshift=-1.0cm, xshift=0.0cm, font=\small] {$\Phi/\Phi_{\rm GR}$};
\end{tikzpicture}
\begin{tikzpicture}
\node (img) {\includegraphics[width=8cm]{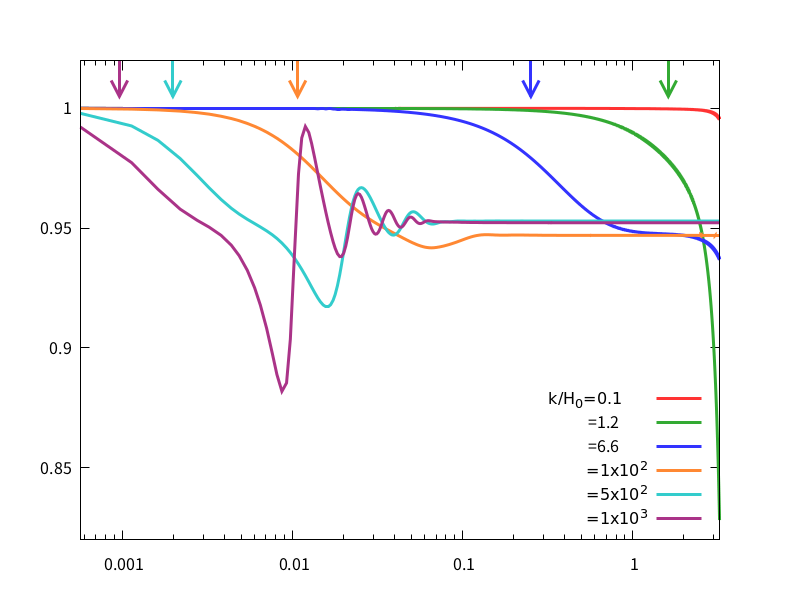}};
\node[below=of img, node distance=0cm, yshift=1.0cm, xshift=0.0cm, anchor=center, font=\small] {$H_0\eta$};
\node[left=of img, node distance=0cm, rotate=90, anchor=center, yshift=-1.0cm, xshift=0.0cm, font=\small] {$\Phi/\Phi_{\rm GR}$};
\end{tikzpicture}
\caption{Time-evolution of $\Phi$ normalized by its GR counterpart for $\beta=0.1$ ({\it left}) and $\beta=1$ ({\it right}).
The arrows hanging on the top horizontal axis indicate the horizon re-entry time for each mode.} 
\label{fig:Phi}
\end{figure}
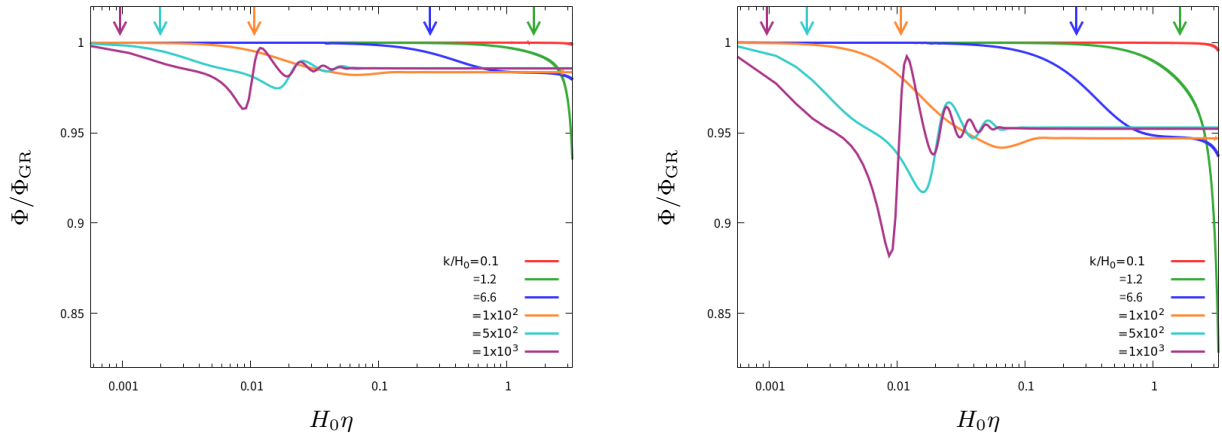

Let us discuss how varying $\beta$
leads to the change in the angular power spectrum as described above.
Figure~\ref{fig:Phi} shows the time evolution of the gravitational potential
$\Phi$ in the modified gravity model divided by that in GR, $\Phi_{\textrm{GR}}$,
for various wavenumbers.
The parameters are given by the ``2DoF'' best-fit ones in
Table \ref{tab:conf_range} except for $\beta$, and
we use $\beta=0.1$ and $\beta=1$ for the left and right panels, respectively, to see how the results depend on $\beta$.
The gravitational potential in GR, $\Phi_{\textrm{GR}}$,
is computed simply by setting $\beta=0$.
(To compute $\Phi_{\textrm{GR}}$
we do not use the ``$\Lambda$CDM'' best-fit parameters in the table.)
It can be seen from these plots that the ratio $\Phi/\Phi_{\textrm{GR}}$ is constant
both in the early and late times. This is the behavior anticipated
from the discussion in the previous section, which validates the argument there.
Setting the initial conditions
so that $\Phi/\Phi_{\textrm{GR}}\to 1$ as $\eta\to 0$,
the gravitational potential at the present time
is reduced by $1\%~(5\%)$ for $\beta=0.1~(1)$
compared to the GR result (see blue, orange, cyan and magenta lines).
The arrows hanging on the top horizontal axis
in Fig.~\ref{fig:Phi} indicate the horizon re-entry time for each mode.
One can see that the potential starts to decay or oscillate around the horizon re-entry time when the deviation of the evolution equations from GR becomes prominent.
Both in the cases with $\beta=0.1$ and $\beta=1$,
the ratio $\Phi/\Phi_{\textrm{GR}}$ with $k/H_0=1.2$, for which horizon re-entry occurs
at a time close to the present time, decays significantly after horizon re-entry.
The same feature is observed for $k/H_0=0.1$ and $6.6$, though the decay is less significant.
This is a transient behavior that occurs around horizon re-entry and
the ratio would settle down to a constant value in the future.

\section{Conclusions}\label{sec:conclusions}

In this paper, we have considered the cosmology of modified gravity
with just two tensorial degrees of freedom and no propagating
scalar mode~\cite{Gao:2019twq}, focusing on a particular subset
of theories which started to be explored in Ref.~\cite{Iyonaga:2021yfv}.
The model is interesting because it evades Solar System tests
and can reproduce exactly the same expansion history of
the $\Lambda$CDM model based on general relativity (GR)~\cite{Iyonaga:2021yfv}.
Moreover, black hole solutions are the same as those in GR and
gravitational waves propagate at the speed of light~\cite{Iyonaga:2021yfv}.
Linear cosmological perturbations are thus the only
way considered so far to discriminate between this modified gravity model
and GR.

To study the evolution of cosmological perturbations
in modified gravity with just two tensorial degrees of freedom,
we have modified the Boltzmann code developed originally for
general scalar-tensor theories~\cite{Hiramatsu:2020fcd}
to implement the present case where the scalar field obeys
a constraint equation rather than a hyperbolic evolution equation.
With this code we have clarified how the modified gravity parameter
$\beta$, which is the only additional parameter of the present model
with respect to the $\Lambda$CDM model, changes
the perturbation evolution and the CMB temperature and E-mode
angular power spectra. We have performed a Markov-Chain Monte-Carlo simulation
to obtain the best-fit cosmological and modified-gravity parameters
from Planck data.
The constraints on the $\beta$ parameter
we have derived read $-0.047<\beta<-0.028$ at 68\% c.l..
This is the first observational test on the
cuscuton-like modified gravity model of Ref.~\cite{Gao:2019twq}
evading other major experimental constraints.

Surprisingly, our result indicates that GR ($\beta=0$) is disfavoured at
$\sim 4\sigma$. The present result, however, is based only on the CMB observation.
Further observational tests need to be promoted to confirm
whether the cuscuton-like modified gravity model is really favoured.
For instance, the large-scale structure survey provides
an independent test for the modified
gravity model through the cosmological evolution of the
matter density fluctuation characterized by $\sigma_8$ or $f\sigma_8$. We leave the joint analysis for future study.

\acknowledgments
The work of TH was supported by JSPS KAKENHI Grant No.~JP21K03559.
The work of TK was supported by
JSPS KAKENHI Grant No.~JP20K03936 and
MEXT-JSPS Grant-in-Aid for Transformative Research Areas (A) ``Extreme Universe'',
No.~JP21H05182 and No.~JP21H05189.


\bibliography{refs}
\bibliographystyle{JHEP}
\end{document}